\documentclass[a4paper]{article}
\RequirePackage[english]{babel}
\RequirePackage[latin1]{inputenc}
\RequirePackage[T1]{fontenc}
\RequirePackage{mathrsfs}
\RequirePackage{amsmath}
\RequirePackage{amssymb}
\RequirePackage{amsbsy}
\RequirePackage{bm}
\begin{document}
\title{\bf{Leptonic Electroweak\\ Spin-Torsion Interactions}}
\author{Luca Fabbri\\ \footnotesize  Dipartimento di Fisica, Universit\`a di Bologna, 
Via Irnerio 46, 40126 Bologna, ITALY}
\date{}
\maketitle
\begin{abstract}
In this paper we consider the most general field equations for a system of two fermions of which one single-handed, showing that the spin-torsion interactions among these spinors have a structure identical to that of the electroweak forces among leptons; possible extensions are discussed.
\end{abstract}
\section*{Introduction}
In the structure of the Dirac field equation given for the most general fermionic dynamics, the most general spinorial derivative contains torsion; since torsion is a tensor then all torsional contribution can be separated away: the most general spinorial derivative with torsion is thus decomposed in terms of the simplest spinorial derivative without torsion plus torsional contributions. Eventually when the field equations coupling torsion to the spin density are used, these torsional contributions get the form of specific spinorial autointeractions \cite{f/0}.

In the case in which many spinors are considered, then the spin density of the system is the sum of all spin densities of each spinor involved; thus the additional interactions display both the form of spinorial autointeraction of the each spinor with itself and spinorial mutual interactions of each spinor with every other reciprocally. These spinorial interactions have a specific structure that looks closely like the form of the leptonic weak forces as in \cite{f/1}.

Therefore in the situation in which we consider a system of two spinors of which one is a spinor with both projections and the other is a semi-spinor with a single left-handed projection, it is possible to prove that the spinorial interactions have a structure that compared to the form of the leptonic weak interactions is identical at all, as we will show here.
\section{Torsional interaction}
In this paper we will employ the notations of \cite{f/0}, only briefly recalling that the most general torsional derivative of spinors is decomposable as
\begin{eqnarray}
&D_{\mu}\psi^{a}=\nabla_{\mu}\psi^{a}+\frac{1}{4}Q_{\mu\nu\sigma}\sigma^{\nu\sigma}\psi^{a}
\label{derivatives}
\end{eqnarray}
in the torsionless simplest derivative of the spinors plus torsional contributions.

A set of $k$ spinor fields labelled with the indices in Latin is governed by a system of spinor field equations
\begin{eqnarray}
&i\gamma^{\mu}D_{\mu}\phi^{a}=0
\label{matter}
\end{eqnarray}
in the massless case, and these equations come along with the background field equations given for the Ricci and Cartan tensors in terms of the energy and the spin density of the spinor field as
\begin{eqnarray}
&G_{\alpha\beta}=\frac{i}{4}\sum_{a}\left[\bar{\psi}^{a}\gamma_{\alpha}D_{\beta}\psi^{a}
-D_{\beta}\bar{\psi}^{a}\gamma_{\alpha}\psi^{a}\right]
\label{metric}
\end{eqnarray}
and
\begin{eqnarray}
&Q_{\mu\alpha\beta}
=-\frac{i}{4}\sum_{a}\bar{\psi}^{a}\{\gamma_{\mu},\sigma_{\alpha\beta}\}\psi^{a}
\label{torsion}
\end{eqnarray}
according to the prescription of the Einstein-Sciama--Kibble scheme; then it is possible to use the field equations (\ref{torsion}) to substitute torsion with the spin density of the spinor fields as
\begin{eqnarray}
&i\gamma^{\mu}\nabla_{\mu}\psi^{a}
+\frac{3}{16}\sum_{b}\bar{\psi}^{b}\gamma_{\mu}\gamma\psi^{b}
\gamma^{\mu}\gamma\psi^{a}=0
\label{matterfield}
\end{eqnarray}
in which now combinations of spinorial bilinear fields have appeared.

These spinorial bilinears give, in terms where $a=b$, autointeractions of the spinor field with itself, and in the terms where $a\neq b$, interactions of the spinor with all other spinor fields that contribute to the spin density; to provide a physical interpretation we may think that as one spinor field $b$ propagates close enough to another spinor field $a$ so that the spin distribution of $b$ has relevant contributions in the spin distribution of $a$, then there is a change in the torsion of the spacetime around $a$ that influences its dynamics. This influence affects the dynamics as an effective interaction, whose structure will be investigated.

\subsection{Torsional interaction:\\ spin coupling of spinor and single-handed spinor}
In this paper we want to compare these spinorial interactions with the leptonic weak forces \cite{f/1}: therefore it is necessary that the two systems have the same spinorial field content, and consequently we are going to consider the case in which two spinor fields are present, and where one is a spinor with both projections while the other is a semi-spinor, with a single left-handed projection. 

In the following we take semi-spinors with single-handed structure as
\begin{eqnarray}
&\bar{\psi}_{L}^{a}\gamma\equiv\bar{\psi}_{L}^{a}\ \ \ \ 
\gamma\psi_{L}^{a}\equiv-\psi_{L}^{a}\ \ \ \ 
\bar{\psi}_{R}^{a}\gamma\equiv-\bar{\psi}_{R}^{a}\ \ \ \ 
\gamma\psi_{R}^{a}\equiv\psi_{R}^{a}
\end{eqnarray}
for the left-handed and the right-handed conjugates spinors respectively.

Moreover we recall that the Fierz rearrangement
\begin{eqnarray}
\nonumber&\psi^{b}\bar{\psi}^{a}\equiv
\frac{1}{4}\bar{\psi}^{a}\psi^{b}\mathbb{I}
-\frac{1}{2}\bar{\psi}^{a}\sigma_{\mu\nu}\psi^{b}\sigma^{\mu\nu}
-\frac{1}{4}i\bar{\psi}^{a}\gamma\psi^{b}i\gamma+\\
&+\frac{1}{4}\bar{\psi}^{a}\gamma_{\mu}\psi^{b}\gamma^{\mu}
-\frac{1}{4}\bar{\psi}^{a}\gamma_{\mu}\gamma\psi^{b}\gamma^{\mu}\gamma
\end{eqnarray}
for any $a$ and $b$ holds identically; in addition we have that
\begin{eqnarray}
\left(\bar{\psi}^{a}\gamma_{\mu}\psi^{a}\gamma^{\mu}
+\bar{\psi}^{a}\gamma_{\mu}\gamma\psi^{a}\gamma^{\mu}\gamma\right)\psi^{a}\equiv0
\end{eqnarray}
with no sum on $a$ is another identity that will be used in the treatment.

In the case we have two spinors the spinor field equations are given by the spinor field equations (\ref{matterfield}) with $k=2$ explicitly written as
\begin{eqnarray}
&i\gamma^{\mu}\nabla_{\mu}\psi^{1}
+\frac{3}{16}\bar{\psi}^{1}\gamma_{\mu}\gamma\psi^{1}\gamma^{\mu}\gamma\psi^{1}
+\frac{3}{16}\bar{\psi}^{2}\gamma_{\mu}\gamma\psi^{2}\gamma^{\mu}\gamma\psi^{1}=0\\
&i\gamma^{\mu}\nabla_{\mu}\psi^{2}
+\frac{3}{16}\bar{\psi}^{1}\gamma_{\mu}\gamma\psi^{1}\gamma^{\mu}\gamma\psi^{2}
+\frac{3}{16}\bar{\psi}^{2}\gamma_{\mu}\gamma\psi^{2}\gamma^{\mu}\gamma\psi^{2}=0
\label{matterfields}
\end{eqnarray}
as it can be seen by separating the fields directly.

In this case we want to deal with a couple of spinors of which one is a spinor with both projections whereas the other is a semi-spinor with a single left-handed projection, and so we separate all right-handed and left-handed projections, and requiring $\psi^{1}_{R}\equiv0$, we have that the field equations above are given by the following
\begin{eqnarray}
\nonumber
&i\gamma^{\mu}\nabla_{\mu}\psi^{1}_{L}
+\frac{3}{16}\bar{\psi}^{1}_{L}\gamma_{\mu}\psi^{1}_{L}\gamma^{\mu}\psi^{1}_{L}
+\frac{3}{16}\bar{\psi}^{2}_{L}\gamma_{\mu}\psi^{2}_{L}\gamma^{\mu}\psi^{1}_{L}-\\
&-\frac{3}{16}\bar{\psi}^{2}_{R}\gamma_{\mu}\psi^{2}_{R}\gamma^{\mu}\psi^{1}_{L}=0\\
\nonumber
&i\gamma^{\mu}\nabla_{\mu}\psi^{2}_{L}
+\frac{3}{16}\bar{\psi}^{1}_{L}\gamma_{\mu}\psi^{1}_{L}\gamma^{\mu}\psi^{2}_{L}
+\frac{3}{16}\bar{\psi}^{2}_{L}\gamma_{\mu}\psi^{2}_{L}\gamma^{\mu}\psi^{2}_{L}-\\
&-\frac{3}{16}\bar{\psi}^{2}_{R}\gamma_{\mu}\psi^{2}_{R}\gamma^{\mu}\psi^{2}_{L}=0\\
\nonumber
&i\gamma^{\mu}\nabla_{\mu}\psi^{2}_{R}
-\frac{3}{16}\bar{\psi}^{1}_{L}\gamma_{\mu}\psi^{1}_{L}\gamma^{\mu}\psi^{2}_{R}
-\frac{3}{16}\bar{\psi}^{2}_{L}\gamma_{\mu}\psi^{2}_{L}\gamma^{\mu}\psi^{2}_{R}+\\
&+\frac{3}{16}\bar{\psi}^{2}_{R}\gamma_{\mu}\psi^{2}_{R}\gamma^{\mu}\psi^{2}_{R}=0
\label{fundamentalmatterfields}
\end{eqnarray}
in which all spinors are either the left-handed or the right-handed semi-spinors written in their irreducible chiral representation.

By using the Fierz rearrangements we can rewrite the field equations in the equivalent form
\begin{eqnarray}
\nonumber
&i\gamma^{\mu}\nabla_{\mu}\psi^{1}_{L}
+\frac{3}{16}\left(\frac{G_{Y}-3}{3}\right)
\left[\frac{1}{2}\left(\bar{\psi}^{1}_{L}\gamma_{\mu}\psi^{1}_{L}
-\bar{\psi}^{2}_{L}\gamma_{\mu}\psi^{2}_{L}\right)\right]\gamma^{\mu}\psi^{1}_{L}+\\
\nonumber
&+\frac{3}{16}\left(\frac{G_{Y}-3}{3}\right)
\left[\bar{\psi}^{2}_{L}\gamma^{\mu}\psi^{1}_{L}\right]\gamma_{\mu}\psi^{2}_{L}+\\
\nonumber
&+\frac{1}{2}\cdot\frac{3}{8}\left(G_{Y}-1\right)\left[\bar{\psi}^{2}_{R}\gamma_{\mu}\psi^{2}_{R}
+\frac{1}{2}\left(\bar{\psi}^{1}_{L}\gamma_{\mu}\psi^{1}_{L}
+\bar{\psi}^{2}_{L}\gamma_{\mu}\psi^{2}_{L}\right)\right]\gamma^{\mu}\psi^{1}_{L}-\\
&-\frac{3}{8}G_{Y}\bar{\psi}^{2}_{R}\psi^{1}_{L}\psi^{2}_{R}=0\\
\nonumber
&i\gamma^{\mu}\nabla_{\mu}\psi^{2}_{L}
+\frac{3}{16}\left(\frac{G_{Y}-3}{3}\right)
\left[\bar{\psi}^{1}_{L}\gamma^{\mu}\psi^{2}_{L}\right]\gamma_{\mu}\psi^{1}_{L}-\\
\nonumber
&-\frac{3}{16}\left(\frac{G_{Y}-3}{3}\right)
\left[\frac{1}{2}\left(\bar{\psi}^{1}_{L}\gamma_{\mu}\psi^{1}_{L}
-\bar{\psi}^{2}_{L}\gamma_{\mu}\psi^{2}_{L}\right)\right]\gamma^{\mu}\psi^{2}_{L}+\\
\nonumber
&+\frac{1}{2}\cdot\frac{3}{8}\left(G_{Y}-1\right)\left[\bar{\psi}^{2}_{R}\gamma_{\mu}\psi^{2}_{R}
+\frac{1}{2}\left(\bar{\psi}^{1}_{L}\gamma_{\mu}\psi^{1}_{L}
+\bar{\psi}^{2}_{L}\gamma_{\mu}\psi^{2}_{L}\right)\right]\gamma^{\mu}\psi^{2}_{L}-\\
&-\frac{3}{8}G_{Y}\bar{\psi}^{2}_{R}\psi^{2}_{L}\psi^{2}_{R}=0\\
\nonumber
&i\gamma^{\mu}\nabla_{\mu}\psi^{2}_{R}
+\frac{3}{8}\left(G_{Y}-1\right)\left[\frac{1}{2}\left(\bar{\psi}^{1}_{L}\gamma_{\mu}\psi^{1}_{L}
+\bar{\psi}^{2}_{L}\gamma_{\mu}\psi^{2}_{L}\right)
+\bar{\psi}^{2}_{R}\gamma_{\mu}\psi^{2}_{R}\right]\gamma^{\mu}\psi^{2}_{R}-\\
&-\frac{3}{8}G_{Y}\left(\bar{\psi}^{1}_{L}\psi^{2}_{R}\psi^{1}_{L}
+\bar{\psi}^{2}_{L}\psi^{2}_{R}\psi^{2}_{L}\right)=0
\label{equivalentfundamentalmatterfields}
\end{eqnarray}
in terms of the parameter $G_{Y}$ which will be useful in following.

In this form the free parameter $G_{Y}$ has been introduced in order to remain in the most general case possible; we notice that the parameter $G_{Y}$ is such that its value determines the sign in front of the interactions: then by choosing for this parameter $G_{Y}$ one of the values given by $1<G_{Y}<3$ we obtain attractive interaction within the field equations. It is possible to see that because the condition $1<G_{Y}<3$ gives attractive interactions then it actually ensures the evolution of these fields to be such that they may form composite scalar bound states, important for what we are going to do next.

\paragraph{Massless fundamental leptons and composite scalar and vector fields.} So far we have started from a system of field equations for two spinors of which one is a semi-spinor with a single left-handed projection writing them in the equivalent form above, which may itself be written now as
\begin{eqnarray}
&i\gamma^{\mu}\nabla_{\mu}L
-\frac{1}{2}g\vec{A}_{\mu}\cdot\vec{\sigma}\gamma^{\mu}L
+\frac{1}{2}g'B_{\mu}\gamma^{\mu}L-G_{Y}\phi R=0\\
&i\gamma^{\mu}\nabla_{\mu}R+g'B_{\mu}\gamma^{\mu}R-G_{Y}\phi^{\dagger}L=0
\label{compactequivalentfundamentalmatterfields}
\end{eqnarray}
which is a form known well. Indeed this is precisely the form of the field equations for the lepton fields before the symmetry breaking in the standard model.

Now in order to better see this we rename the couple of spinor fields
\begin{eqnarray}
&\left(\psi^{2}_{R}\right)=R\ \ \ \ \ \ \ \ 
\left(\begin{tabular}{c}
$\psi^{1}_{L}$\\ $\psi^{2}_{L}$
\end{tabular}\right)=L
\end{eqnarray}
as new lepton fields: then considering their bilinear fields we also define
\begin{eqnarray}
&\frac{3}{8}\bar{R}L=\phi
\end{eqnarray}
for the scalar field; and finally we define
\begin{eqnarray}
&\frac{3}{8}\left(\bar{L}\gamma_{\mu}\frac{\mathbb{I}}{2}L+\bar{R}\gamma_{\mu}R\right)
=-\left(\frac{g'}{1-G_{Y}}\right)B_{\mu}\\
&\frac{3}{8}\bar{L}\gamma_{\mu}\frac{\vec{\sigma}}{2}L=\left(\frac{3g}{3-G_{Y}}\right)\vec{A}_{\mu}
\end{eqnarray}
for the vector fields. We see that from the fundamental lepton fields it is possible to build a complex singlet and a complex doublet: from these we also see that the composite scalar field is a complex doublet; and the composite vector fields are a real singlet and a real triplet. In fact, consider that the massless fundamental leptons $R$ and $L$ are complex and therefore they can be subject to two independent $U(1)$ transformations: the massless composite scalar field given by the definition $\frac{3}{8}\bar{R}L$ is consequently complex and thus it transforms according to the combined $U(1)$ transformations; finally the massless composite vector fields are real and do not transform at all. Moreover we have that the massless fundamental leptons $R$ and $L$ are such that $R$ is the only right-handed spinor and it does not transform into anything else whereas $L$ is formed by a doublet of left-handed spinors which in principle are indistinguishable and therefore they can transform into one another according to an $SU(2)_{L}$ transformation: the massless composite scalar field $\frac{3}{8}\bar{R}L$ transforms according to the $SU(2)_{L}$ transformation above; finally the massless composite vector fields are such that the vector field given by the right-handed projections $\bar{R}\gamma_{\mu}R$ does not transform whereas on the other hand the vector fields given by the left-handed projections are such that the vector field given by $\bar{L}\gamma_{\mu}L$ transforms into itself while the three vector fields $\bar{L}\gamma_{\mu}\vec{\sigma}L$ transform into each other and so the vector fields given by $2\bar{R}\gamma_{\mu}R+\bar{L}\gamma_{\mu}L$ and $\bar{L}\gamma_{\mu}\vec{\sigma}L$ transform as the singlet and the triplet of the adjoint representation of the same $SU(2)_{L}$ transformation that has been given above. This establishes the transformation laws of the lepton, the scalar and vector fields before the symmetry breaking in the standard model.

That the torsional interactions and the weak forces may be linked has been conjectured long ago \cite{s-s,h-h-k-n,s-s/1,h-d,s-g,s-s/2}, but that torsional and weak field could be connected so tightly to make us think they could even be the same field is a thesis that has never been discussed deeply in the literature before.

\subparagraph{Massless fundamental leptons and composite scalar and vector fields: structure of $U(1)\times SU(2)_{L}$ local electroweak gauge interaction.} Finally we have that the system of field equations (\ref{compactequivalentfundamentalmatterfields}) can be written as
\begin{eqnarray}
&i\gamma^{\mu}\mathbb{D}_{\mu}L-G_{Y}\phi R=0\\
&i\gamma^{\mu}\mathbb{D}_{\mu}R-G_{Y}\phi^{\dagger}L=0
\label{invariantcompactequivalentfundamentalmatterfields}
\end{eqnarray}
in which the derivatives have been written in a compact form.

This compact form is obtained upon definition of the derivatives
\begin{eqnarray}
&\mathbb{D}_{\mu}L=\nabla_{\mu}L
+\frac{i}{2}(g\vec{A}_{\mu}\cdot\vec{\sigma}-g'B_{\mu})L\\
&\mathbb{D}_{\mu}R=\nabla_{\mu}R-ig'B_{\mu}R
\end{eqnarray}
being covariant for general $U(1)\times SU(2)_{L}$ local transformations. This generalization is possible since the massless fundamental leptons $R$ and $L$ are functions of the spacetime position and so their mixing may take place with coefficients depending on the spacetime position themselves. This defines the covariant derivatives of lepton fields before the symmetry breaking in the standard model.

As the spinor field equations coupled together describe fields that are massless while being in interactions with one another, then solutions may be found in the particular form given by
\begin{eqnarray}
&i\mathbb{D}_{\mu}L=q_{\mu}L+\frac{1}{4}G_{Y}\gamma_{\mu}\phi R
\ \ \ \ \ \ \ \ q_{\mu}\gamma^{\mu}L=0\\
&i\mathbb{D}_{\mu}R=p_{\mu}R+\frac{1}{4}G_{Y}\gamma_{\mu}\phi^{\dagger}L
\ \ \ \ \ \ \ \ p_{\mu}\gamma^{\mu}R=0
\label{solutions}
\end{eqnarray}
where left-handed and right-handed fields have momenta $q_{\mu}$ and $p_{\mu}$ in leading order and in which the interactions are in the following order of approximation.

We recall that because the spinorial solutions have behaviour that creates the repulsive effects of the exclusion principle then spinor fields cannot condensate into scalar bound states; therefore to overcome this circumstance it is necessary to assume for spinor field solutions conditions for which they turn out to be able to form compound scalar bound states: it is in the situation for which the condition given by $1<G_{Y}<3$ above holds that composite scalar bound states form. And henceforth the condition $1<G_{Y}<3$ is what consequently gives the possibility to have a stable scalar field as the one we have defined here.

Now because it is in terms of the massless fundamental leptons that we build the massless composite scalar field and since for the massless fundamental leptons we have the spinor field equations (\ref{invariantcompactequivalentfundamentalmatterfields}) then it is possible for the massless composite scalar field to derive the field equations given by 
\begin{eqnarray}
\nonumber
&\mathbb{D}^{2}\phi+2G_{Y}^{2}\phi^{2}\phi-\frac{3}{4}\mathbb{D}_{\mu}\bar{R}\mathbb{D}^{\mu}L
-\frac{3}{8}iG_{Y}
\left(\bar{L}\gamma^{\mu}\mathbb{D}_{\mu}\phi L-\bar{R}\gamma^{\mu}\mathbb{D}_{\mu}\phi R\right)+\\
&+\frac{3i}{16}
\left[g(\nabla_{[\mu}\vec{A}_{\nu]}-g\vec{A}_{\mu}\times\vec{A}_{\nu})\cdot\vec{\sigma}
-3g'\nabla_{[\mu}B_{\nu]}\right]\bar{R}\sigma^{\mu\nu}L=0
\end{eqnarray}
covariant under the general $U(1)\times SU(2)_{L}$ local transformations above.

By employing the solutions (\ref{solutions}) we have that the massless composite scalar field is ruled by the equation given by
\begin{eqnarray}
&\mathbb{D}^{2}\phi
+\frac{G_{Y}^{2}}{2}[\phi^{2}-\frac{4}{G_{Y}}(\frac{4}{3}+\frac{p_{\mu}q^{\mu}}{G_{Y}})]\phi
+G_{Y}\bar{R}L=0
\label{scalarfields}
\end{eqnarray}
in terms of the scalar product of the momenta $q_{\mu}p^{\mu}$ in leading order and with interactions in the following order of approximation therefore neglected: with this form for the scalar field equation the stable vacuum configuration is assumed for solutions verifying the constraint
\begin{eqnarray}
&\phi^{2}=\frac{4}{G_{Y}}(\frac{4}{3}+\frac{p_{\mu}q^{\mu}}{G_{Y}})
\label{stablevacuumsolutions}
\end{eqnarray}
in terms of the scalar product between the momenta $p_{\mu}q^{\mu}$ so that the vacuum expectation value is subject to a scaling that depends on the energy squared and we remind that once again all interactions in the higher-order of approximation have been neglected. In this way we have that there is spontaneous breakdown of the symmetry after which the mechanism of mass generation assigning to the lepton field a mass that is equal to the scalar mass takes place.

Some comments are due: in the first place, the mass of the scalar field depends on the coupling constant $G_{Y}$ and the energy scale $p_{\mu}q^{\mu}$, and ultimately on the mass of the lepton we are considering; then we know that none of the known leptons possesses a mass that is high enough to give to the scalar field a mass larger than the Linde-Weinberg lower bound \cite{l,w}, with the consequence that the contribution of the effective potential determining the spontaneous breakdown of the symmetry are not larger than the contributions coming from radiative corrections, and the stability of the vacuum configuration is compromised. Thus the case we have presented here does not reproduce the correct dynamics for the scalar field if we want it to be the scalar field responsible for the spontaneous breaking of the symmetry and the mass generation; on the other hand however we have only considered lepton families and obviously more complete cases in which also hadron families are included will certainly give richer dynamics. Maybe in this case we can find more general mass relations assigning the correct mass to the composite scalar field that should eventually correspond to the Higgs field of the standard model.
\section*{Conclusion}
In this paper we have proved that the field equations for a spinor and a semi-spinor of left-handed helicity coupled through torsion and free of any other interaction are formally equivalent to the field equations for massless leptons without torsion but with electroweak gauge interactions; to proceed in logical order next step would be to know whether this derivation can be extended to more general systems of field equations for two spinors coupled through torsion and free of any other interaction to see whether they are formally equivalent to the field equations of massless hadrons without torsion but with the electroweak gauge interactions: if this result is obtained then it would mean that the field equations with torsion in the free case are formally equivalent to field equations without torsion but with the electroweak gauge interactions. This would eventually mean that within the field equations for the fermion fields the presence of the torsion would be manifest in the guise of the electroweak gauge interactions.

So starting from field equations for massless spinors it is possible to see that the torsional interactions are formally equivalent to electroweak gauge interactions, all this in the situation of masslessness for fermions and therefore when the gauge symmetry is still intact: then gauge symmetry breaking for the mass generation is given in terms of the Higgs field here as in the standard model, with the difference that it is in terms of composite bound states of fermions that the Higgs field is given here whereas it is in terms of fundamental degrees of freedom that it is given in the standard model. This fact is not so surprising if we think that the field equations from which we started are also equivalent to those of the Nambu-Jona--Lasinio model, for which a mechanism of chiral symmetry breaking for the mass generation is discussed in \cite{n-j--l}.

As an additional comment, we remark that since the torsion is thought to be the potential of covariant spacetime translations whereas the gauge interactions arise from the gauging of internal transformations then the relation between torsion and gauge fields may be mirrored into the link between spacetime translations and internal transformations: we can speculate that such a link might be established under the hypotheses of the Coleman-Mandula theorem \cite{c-m}.

Among the open problems that may affect this model is the fact that here the torsional interaction has mass dimension that is higher than the one required for it to be renormalizable; on the other hand, non-renormalizabillity does not mean that a real ultraviolet divergence has place but only that new effects must enter at high energies: these new effects are likely to be given by the presence of the torsion tensor. For a discussion on this subject see also \cite{p}.

A final open problem is related to the fact that the present torsional interaction is supposed to become relevant at the Planck scale while for our approach to be fully satisfactory it must be relevant already at the Fermi scale: hypotheses based on double typical scales have been discussed in \cite{z}.

These two problems may have a single solution relating different strengths to different scales by an energy-dependent coupling, but we ignore if such gravitational running coupling is feasible.


\begin{thebibliography}{25}
\bibitem{f/0}
L.~Fabbri,
arXiv:1006.3672 [gr-qc].
\bibitem{f/1}
L.~Fabbri,
arXiv:0908.2349 [hep-th].
\bibitem{s-s}
V. De Sabbata and C.~Sivaram, ``Spin and Torsion in Gravitation''\\
(World Scientific, Singapore, 1994).
\bibitem{h-h-k-n}
F.~W.~Hehl, P.~Von Der Heyde, G.~D.~Kerlick and J.~M.~Nester,\\
\textit{Rev. Mod. Phys.} \textbf{48}, 393 (1976).
\bibitem{s-s/1}
C.~Sivaram and K.~P.~Sinha,
\textit{Phys. Rept.} \textbf{51}, 111 (1979).
\bibitem{h-d} 
F.~W.~Hehl and B.~K.~Datta, 
\textit{J. Math. Phys.} \textbf{12}, 1334 (1971).
\bibitem{s-g}
V.~De Sabbata and M.~Gasperini,
\textit{Gen. Rel. Grav.} \textbf{10}, 731 (1979).
\bibitem{s-s/2}
C.~Sivaram and K.~P.~Sinha,
\textit{Lett. Nuovo Cim.} \textbf{13}, 357 (1975).
\bibitem{l}
A.~Linde,
\textit{Sov. Phys. JETP} \textbf{23}, 64 (1976).
\bibitem{w}
S.~Weinberg,
\textit{Phys. Rev. Lett.} \textbf{36}, 294 (1976).
\bibitem{n-j--l}
Y.~Nambu and G.~Jona--Lasinio,
\textit{Phys. Rev.} \textbf{124}, 246 (1961).
\bibitem{c-m}
S.~Coleman and J.~Mandula,
\textit{Phys. Rev.} \textbf{159}, 1251 (1967).
\bibitem{p}
N.~J.~Poplawski,
\textit{Phys. Lett.} \textbf{B690}, 73 (2010).
\bibitem{z}
M.~A.~Zubkov,
\textit{Mod. Phys. Lett. A} \textbf{25}, 2885 (2010).
\end{thebibliography}
\end{document}